# Optimizing the Replay Protection at the Link Layer Security Framework in Wireless Sensor Networks

Devesh C. Jinwala, Dhiren R. Patel, Members, IAENG, Sankita Patel and Kankar S. Dasgupta

*Abstract*—Ensuring communications security in Wireless Sensor Networks (WSNs) is very vital because the security protocols therein, should be devised to work at the link layer. Theoretically, any link layer security protocol must support three vital security attributes viz. Confidentiality, Message Integrity and Replay protection.

However, in order to ensure lesser overhead, replay protection is often not incorporated as part of the link layer security framework. We argue here, that it is essential to implement replay protection at the link layer only and devise a simple scheme to do so. We first survey the common approaches to ensuring replay protection in conventional networks. We also implement the conventional algorithms for replay protection using the link layer framework for WSNs viz. TinySec as the underlying platform. Subsequently analyzing their limitations, we propose a novel Bloom-filter based replay protection algorithm for unicast communications. We show that our algorithm is better than the other contemporary approaches for ensuring replay protection in unicast communications in the WSNs.

## I. INTRODUCTION

The Wireless Sensor Networks (WSNs) consist of the sensor nodes collaborating with each other to achieve a specific purpose. There are numerous applications for which the WSNs can potentially be employed. At the same time, the security and privacy issues in WSNs demand a critical examination. This is primarily due to the severe resource constraints that the basic building blocks of the WSNs viz. the sensor nodes, suffer from and the inherent higher resource demands that the conventional security algorithms exhibit. In addition, with the wireless, *open-to-all* communication paradigm followed by WSNs, this problem assumes larger proportions [1].

Moreover, the WSNs follow the *in-network* processing oriented *data-centric* multihop communication, necessitating the security support therein, to be devised at the *link layer*, thereby significantly increasing the overall overhead [2]. The necessity of link layer security support also obviates the direct application of the conventional *header-centric* end-to-end security protocols viz. SSL[3], IPSec[4] in WSNs..

The security attributes desired in a link layer security architecture in WSNs are *data confidentiality* (mechanism: data encryption) and *data integrity* with *data-origin authentication* (mechanism: keyed hash functions) and *message freshness* with protection against the replayed packets.

However, there is no unanimity on whether replay protection should be implemented at the link layer or at the application layer in WSNs. This is exemplified by the fact that the popular link layer security architecture viz. TinySec [5] and the link layer security architecture for body networks viz. SenSec [6] leave replay protection to be supported at the application layer. Both these architectures support security in unicast setup and use the Mica2 motes [7]. MiniSec [8] on the other hand implements replay protection at the link layer. A recent attempt at devising the link layer security architecture for the Moteiv's Telos motes [9], MiniSec employs two different approaches to support replay protection viz. one to work in unicast mode and the other to work in the broadcast mode.

Here, we argue first that the replay protection must indeed be implemented at the link layer and subsequently demonstrate that the approach that MiniSec employs for ensuring replay protection, in unicast setup viz. the *Counter-based* approach, suffers from scalability and synchronization concerns. Therefore, we attempt to explore the alternatives for providing suitable anti-replay protection schemes and propose a simpler Bloom filter based scheme for the same.

In this paper our basic aims are

(a) to justify the case for devising replay protection at the link layer in WSNs.

(b) to explore and describe the conventional techniques for ensuring replay protection.

(c) to adapt the conventional techniques for WSNs and implement them in the link layer framework.

(d) to present a sophisticated Bloom filter [10] based replay protection scheme in unicast setup and,

(e) to evaluate our scheme against the conventional techniques for doing so and justify analytically as well as

experimentally, the advantages in our algorithm.

Our experimental results clearly demonstrate that the proposed Bloom filter based algorithm works well, adding less than 10% of overhead in memory and 0.2% in energy, as compared to the one without replay protection. We also demonstrate that the proposed scheme in unicast setup is simpler than the one employed in MiniSec for broadcast communication.

The rest of the paper is organized as follows: in section 2, we discuss the necessary theoretical background describing the link layer security issues in the WSNs. In section 3, we present a survey of the conventional approaches for ensuring replay protection. In section 4, the approaches for replay protection in WSNs and the Bloom filter based approach for replay protection are discussed along with our design. In section 5, we discuss the methodology and tools employed for implementation whereas in section 6, we demonstrate and analyze the experimental results for various algorithms with respect to storage, computational and energy overhead. Finally, we conclude with the impact of our work and enlist the possibilities of future extensions and applications.

## II. SECURITY IN WIRELESS SENSOR NETWORKS

In this section, we discuss the need for devising the security support at the link layer in WSNs and enlist the required security attributes.

### A. Communication Paradigm in WSNs

The WSNs follow *data-centric* multihop communication paradigm instead of the conventional *header-centric* multihop communication. The data-centric multihop communication relies on the *in-network* processing of the data. In-network processing is the processing of the sensed data, done on-the-fly, during the transmission of the data to the base station [5]. It consists of trivial but significant operations like *summarization, duplicate elimination* or *aggregation* of the data, done by the intermediate sensor nodes themselves, to reduce the total number of packets actually communicated to the base station. The principal advantage of such in-network processing is the overall reduced communication costs [5].

However, due to in-network processing, the conventional *end-to-end* security (e.g. SSL, IPSec) protocols, that do not compel the intermediate network nodes to observe the contents of the data packets; cannot be used in WSNs. Thus, with the need to investigate the data packets at each hop (data-centric multihop communication), the security support has to be implemented at the link layer.

However, due to the *link layer* security support, the overall resource overhead that has anyway increased due to the addition of security mechanisms, further increases. Therefore, it is essential to tune carefully, the security protocols for the link layer security architecture for WSNs.

There indeed have been a significant number of attempts published in the literature, aimed at enriching the link layer protocol with the essential security mechanisms. We present characteristic details of such protocols later. However for now, we describe the principal goals of any link layer security architecture.

### B. Goals of the link layer architecture

The security attributes desired from a link layer security solution for WSNs are

(a) *Confidentiality:* Confidentiality of the data is achieved by encrypting it using a symmetric key block cipher. The data encryption can be achieved with a shared key *k* and the functions $E_k(m)$ and $D_k(m)$ for encryption and decryption respectively. That is, for the plain text *m*, the cipher text is given by $c=E_k(m)$, whereas the plaintext is obtained by $m=D_k(c)$. For such a symmetric key scheme, it is quite clear that the sender and the receiver must first secretly choose or exchange the key *k*, to be used subsequently.

(b) *Data integrity:* Data integrity ensures that the data is not altered maliciously, during its transmission from the source to the destination. Typically, the un-keyed hash functions can be employed to ensure the data integrity.

However, along with the data integrity, it is essential to ensure *data origin* (entity) authentication. To achieve the data integrity as well as entity authentication, a keyed hash function, known as the Message Authentication Code (MAC) algorithm is employed.

A MAC algorithm is typically a family of functions $h_k(k, x)$ that takes a secret key and the message as input and generates a MAC-value. If M is the input message, *C* is the MAC algorithm, *K* is the shared secret key and MAC is the message authentication code generated, then $MAC = C_K(M)$.

The sender upon generating the MAC would send the message as

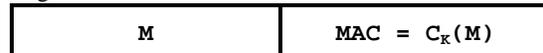

Fig. 1 Generic MAC packet format

The recipient upon receiving the message would use the same function $C_K$ with the same shared secret key K to compute the MAC value. If the computed MAC value and the received MAC value match, the integrity is verified otherwise, it is not.

(c) *Replay Protection:* Replay protection ensures security against replay attacks. A replay attack is defined as *an attack that is carried out on a security protocol by sending a packet that belongs to an earlier context, in the context that is current and under consideration*. Since the packet is not modified in any way, the message integrity checks done by the recipient do not reveal that the packet is actually a replayed packet and not the genuine one. We formally present a discussion of replay attacks and the means of countering them in the next section.

(d) *Message freshness:* The message freshness parameter captures the gap between the transmission of data from the source and its delivery to the recipient. Message freshness can be either *Strong* freshness depicting a measure of *time synchronization* between the communicating entities or *Weak* freshness depicting only *ordering* of the communication undertaken. Message replay protection is a special case of ensuring message freshness.

(e) *Availability:* The security properties to be satisfied must not prevent the message from being available in time, to the recipient. For any security protocol, availability of the data to infer vital information is always of prime concern.

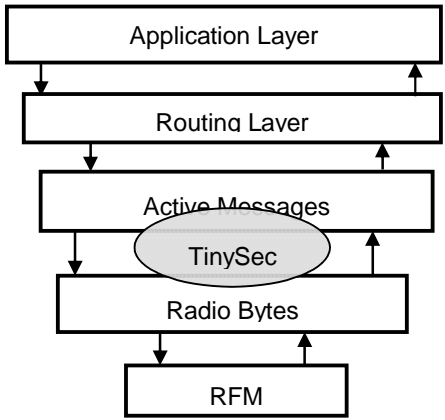

Fig. 2 TinySec in TinyOS protocol stack

(f) *Low Overhead:* The *resource overhead* concerns with the overall expenditure in *computational, storage* and *energy* resources, due to the implementation of the security mechanisms. Naturally the algorithms used to implement the security properties do indeed entail some additional cost. In the resource constrained WSN environment, it is essential to ensure that this overhead is at the minimum, in order to optimize the cost/benefit ratio.

Having discussed the goals of the link layer security architecture, we now survey the existing implemented versions of the same.

*C. Current Link Layer Security Architectures*

The implementation platform that we use for proposing and evaluating the replay protection schemes is TinySec. Karlof *et al* in [5] present TinySec as a lightweight and efficient link-layer security protocol for WSNs. TinySec attempts to provide configurable link layer security with three different modes of operations, providing either (a) no security support or (b) support for message authentication only based on Cipher Block Chaining Message Authentication Code (CBC-MAC) [11] or (c) support for (b) and message confidentiality via encryption in Cipher Block Chaining (CBC) [12] mode.

In the TinyOS protocol stack, TinySec can be viewed as shown in Fig. 2. The TinySec protocol sits in between the Radio bytes layer and the Active Message (AM) layer of the operating system. Thus, whenever an AM packet is required to be transmitted securely, TinySec encrypts the data payload and generates a MAC.

The MAC is generated over the encrypted data and the packet header and is transmitted with the packet. In case the security mode selected is message authentication only, the MAC is generated over the data payload and the packet header, without packet encryption.

The modified packet is sent bit by bit by the radio component. When received at the other end, the bits pass through the RFM on to the Radio bytes layer and then authenticated and decrypted before being passed to the application layer.

TinySec employs 80-bit key-sized Skipjack [13], as the default block cipher. We have used TinySec for our experimentation, because it is open-source and follows a modular plug-in oriented design so as to enable further experimentations. We must emphasize that our results are independent of the link layer protocol implementation i.e. we expect similar results when using any other protocol.

SenSec [6] is another attempt at designing the link layer security framework for specific body monitoring application at the Institute of Infocomm Research, Singapore. Being designed for a specific application, it cannot be employed as a platform for further cryptographic experimentations. It also uses Skipjack as the block cipher with only 80-bit key size but with modified block cipher mode of operation. SenSec also does not support link layer replay protection.

TinySec and SenSec are devised for security support in only unicast communication mode.

MiniSec [8] is another recent attempt at designing the link layer security architecture. MiniSec also uses Skipjack as the block cipher. As mentioned earlier, MiniSec employs two different modes of operation to support the unicast and broadcast communications in WSNs.

III. REPLAY PROTECTION: CONVENTIONAL APPROACHES

Formally, let a sender $A$, at time instant $t_1$ has sent $k$ packets in the set $T$ bearing sequence numbers viz.

$$T_{SEQ} = \{SEQ_1, SEQ_2, SEQ_3, \ldots SEQ_k\} \quad (1)$$

Let $m$ be the number of packets that have been received by the receiver $B$. Due to the packet loss and out of order delivery, $B$ might not receive all the packets that are transmitted. However, let

$$R = \{r_2, r_3, r_6, r_9 \ldots r_m\} \quad (2)$$

be the packets received as per their order of arrival by the receiver $B$ with their sequence numbers given by

$$R_{SEQ} = \{SEQ_2, SEQ_3, SEQ_6, SEQ_9, \ldots SEQ_m\} \; (m<k) \quad (3)$$

To orchestrate a replay attack, the adversary would store any of the packets in the set of received packets $R$ and replay them later.

Replay attacks have been elaborately studied in the conventional networks. In [14], Paul Syverson gives a detailed taxonomy of replay attacks, including the *type* and the *level* of the replay attacks. The focus in this paper is not on analysis of the techniques for preventing replay attacks but specifically on presenting a classification of attacks that is independent of the methods to tackle replay attacks. Syverson fundamentally classifies the replay attacks in terms of two categories viz. the replay attack on the protocol run at the source and that at the destination. The former could be either replay of messages from outside the current run of the protocol or its counterpart, whereas, the latter could be either *deflection* i.e. message being directed to other than the intended recipient, or *reflection* i.e. message being replayed back to the origin or *straight replay* i.e. message being replayed to the intended recipient.

The theoretical solution for providing the replay protection is based on defining a *temporal* or *causal* relationship, between the properties of the message received at an instant and those of the messages received previously. Such relationship is required to be exploited to define an anti-replay protocol to handle a replay attack.

In order to establish such relationship, some information about the message is required to be tagged with it, when the message is transmitted. In [15], U. Carlsen describes the types

```
Algorithm
CounterReplayDetect(CounterReceived,
                               NodeID)
{
1.   id = o;
2.   for id = 1 to lastValidId {
3.     if (id==NodeID) {
4.       if (CounterReceived <=
                 LastCount[NodeID])
5.         replayed = 1;
           else
6.         replayed = 0;
7.     LastCount[NodeID]=CounterReceived;
       }
     }
}
```

Fig. 3 Antireplay Algorithm `CounterReplayDetect`

of information that can be *tagged* with the message viz. any one or all of *protocol identifier, protocol run identifier, primitive types of data items, transmission step identifier, message component identifier*.

In fact, there are wide and diverse approaches as regards to what information can be used as the basis to detect replay. T.Y.C. Woo and S.S. Lam in [16] propose the *principle of full information, i.e.* tagging all the information available with the sender, at the time of transmission of the message. Conversely, Aura in [17] proposes tagging only *a hash* of part of some data that is already known to the recipient.

Nevertheless, irrespective of the information tagged with the message, the anti-replay protocol must let the recipient determine whether the message received is a replayed message or not. As for example, when employing a *transmission step* identifier i.e. a *counter*, the sender tags incremental values of sequence numbers of the messages to each message transmitted. The protocol on the recipient node uses the received values of sequence numbers to determine whether the packet just received is a replayed or not. We formalize the algorithm that employs this method in Fig. 3.

This algorithm does not handle the packets received out of order at the recipient. To handle the same, it can be augmented with a sliding window that checks for the counter value within a defined window.

When employing such counter based method in a network of *n* nodes, each node will have to maintain counter values for *(n-1)* other nodes and hence, if *B* bytes are required for one value of the counter, the total storage expended in all the nodes is

$$\text{StorageOverhead}_{\text{CounterReplayDetect}} = B*(n(n-1)/2) \quad (4)$$

Thus, if the network consists of only 50 nodes and two bytes are used for *counter*, then the total bytes expended is 2450 bytes. This is approximately 5% of the total available RAM in Mica2 motes; implying that 5% of the total RAM is used only for handling *one* security attribute, irrespective of the *cipher*, the *key-size* and the *MAC-size* employed. Hence, there is need for exploring alternative techniques to ensure replay protection.

Alternatively, other possible type tag identifier is a employing a time-stamp. This method is based on a local timestamp say $\{T_x\}_K$ being affixed to the packet, by the sender *X,* sharing a symmetric key *K* with the recipient, at the time of transmission. The recipient node *Y* would decrypt the packet using the shared key *K* and compare the timestamp with its own local clock $T_y$. Based on the comparison, Y would treat the received packet as a fresh packet, only if according to its own judgment, the time drift $|T_y - T_x|$ is sufficiently small.

The issues associated with this approach are that the clock maintained by all the network nodes must be synchronized and the clock drifts amongst the nodes has to be accounted for by the recipient replay detection protocol.

Similar strategies to deal with the replay attacks in conventional networks have indeed been studied and analyzed significantly.

In [18], Li Gong presents a discussion on the choice of identifying the freshness identifier, to be employed with the message. Gong analyzes the potential complexity in devising an anti replay algorithm based on the selection of the freshness identifiers.

In [19], Sreekanth Malladi, Jim Alves-foss and Robert B. Heckendorn present a simple protocol model for preventing replay attacks. They specifically address the issue of identifying the type tags to be attached to the message so as to identify a replayed packet.

In [17], T. Aura discusses in general, a set of design principles that can be used to avoid replay attacks. In this attempt too, the focus is on identifying how to type-tag messages – including the *Principle of Full information* as referenced earlier - to ensure the detection of replay. Though a dated attempt, the author emphasizes that the design principles can be easily applied to real protocols, too.

A typical form of replay attack viz. *session replay* is discussed in [20]. A session replay is one, wherein the login ID and the password of genuine users, are captured to be re-used maliciously later by the adversary, to gain unauthorized access. The authors also discuss simple solution viz. entity authentication based on the One-Time-Password (OTP), to prevent such session replay attacks.

Having surveyed the conventional approaches to anti-replay, we now focus on the same in WSNs.

IV. REPLAY PROTECTION IN WIRELESS SENSOR NETWORKS

In this section, we first emphasize why is it worth investing in the extra resources, for ensuring replay protection in the resource starved sensor nodes; even though the temptation could be to omit it, because of the paucity of resources. We also show why replay protection should be implemented at the link layer only in WSNs. Subsequently we survey the current approaches to anti-replay in WSNs. Later in section IV(C), we discuss the data structure used by us for replay protection.

*A. Motivation for WSNs*

Apart from the confidentiality and integrity (message & entity), in implementing the third security attribute for a link layer security architecture for WSNs, the protocol designer has three design alternatives available:
(a)  Case 1: devising no anti-replay algorithm
(b) Case 2: implementing anti-replay algorithms to work at the application layer and not at the link layer
(c)  Case 3: implementing such a algorithm within the link layer security protocol as one of its implicit goals.

We now consider the pros and cons of all the three cases.

Case 1: If due to the scarcity of overall resources in WSNs, anti-replay schemes are omitted, then, an adversary can cleverly exploit the lack of it, by replaying a genuine packet. For such replayed packets too, the recipient node would still carry out the data integrity check (e.g. using CBC-MAC [11] to verify the MAC) and then decrypt the valid packet. Upon decryption, the recipient node may or may not be able to identify the packet as the replayed packet.

In case it identifies the packet as a replayed packet from the semantics of the data (which may not be the case in every occurrence of such event), it would have anyway wasted all of its precious resources thus far, in futile decryption and integrity checks on the received packet.

In case, it is not able to detect it as a replayed packet then it can be every more serious proposition, as it would still waste its resources in further processing of such message.

In either case, such wastage of resources should essentially be prevented in the resource starved sensor nodes.

Case 2: On the other hand, if the replay protection is ensured only at the application layer, then, with the data-centric multihop communication, the victim WSN may have routed, a packet replayed by an adversary, many hops, before it is detected. This kind of attack will again, waste precious energy and bandwidth resources in the targeted nodes.

Case 3: As compared, if the link layer security architecture were designed with anti-replay protection built-in, it would be possible to detect a replayed packet, when it is first injected into the network; thereby saving not only the computational and storage resources; but also the precious energy.

Thus, it is essential to devise anti-replay protection in the link layer security protocols only. We thus show that our experimental evaluations are indeed worth the efforts put in, to do so.

*B. Replay protection in WSNs*

The principal attempts at devising link layer security framework in the WSNs are the SPINS [21], TinySec [5], SenSec [6] and MiniSec [8].

SNEP as part of SPINS is one of the first attempts at implementing a secure link layer protocol. It achieves replay protection by keeping a consistent counter between the sender and receiver. However, as was discussed in section III earlier and as our experimental results testify, a counter based approach is grossly inefficient in WSNs.

In TinySec, the designers make an attempt at providing configurable security, by providing three different modes of operation wherein

(a) either the application can be compiled to offer both the confidentiality and message integrity viz. *TinySec_AE* mode.

(b) or provide only message integrity viz. *TinySec_Auth* mode.

(c) or do not enable any security features.

However, as mentioned before, the designers of TinySec argue against devising any replay protection protocol at the link layer. We already disproved their argument before.

MiniSec is a secure link layer protocol that serves to for provide, according to its designers, low energy consumption and high security. MiniSec provides two different strategies for anti-replay for its two different operating modes.

For the unicast mode, a *synchronized counter* based approach is proposed. However, as we show in the next section, such scheme is not scalable with the increase in the number of nodes in the network. In addition, it requires costly *resynchronization* routines to be executed, when the counters shared, are desynchronized due to the out-of-order delivery of the packets. As against that, our sliding window based counter approach does not require any resynchronization to be done. However, our counter-based algorithm is not scalable, as in MiniSec.

To handle scalability concerns, MiniSec indeed employs a Bloom Filter based approach with sliding window (in broadcast setup). However, since our focus is only on unicast communication, we consider and compare our results against only the unicast mode of MiniSec.

*C. The Bloom Filter*

Bloom filter is a space-efficient data structure used for fast probabilistic membership tests. It is a way of employing a number of hash functions and to use their output, to determine the set membership on large data sets, efficiently [24].

Formally, a Bloom filter is a vector of *n* bits consisting of each individually addressable cells viz. $a_1, a_2, a_3, a_4.....a_n$ along with *m* different hash functions viz. $h_1, h_2, h_3,..........h_m$. Initially, the Bloom filter is empty with all its bits set to 0.

To add a data element, say $d_i$, to the Bloom filter, each of the hash functions $h_j$, $1<=j<=m$ is applied on $d_i$ to get the hash values $hv_j$. The vector addresses indexed by $hv_j$ in the bloom filter are then set to 1.

Subsequently, to query for the membership of a data element $d_i$, the data element is again hashed with *m* hash functions to again get the *m* different hash values viz. $hv_1, hv_2, hv_3,..........hv_m$. If the bit at any of these $hv_j$ locations in the vector is 0, the element is considered to be not in the set. For, if the element were in the set, then all the bits corresponding to $hv_j$ would have been set to 1 when the element was initially inserted.

Here, it is possible that the bit corresponding to a data element in the vector has been set to 1, during the insertion of other elements. This situation indicates a *false positive* i.e. an element being flagged as being in the set, even if actually it is not.

On the other hand, a false negative i.e. the case where a query returns false, when the element is in fact a member of the set, is guaranteed to be not possible.

Now, this same approach employed for set membership tests can be used for detection of replayed packets too. In such a scenario, when a packet is first received by a node, the message tag in the packet is hashed by *m* hash functions under consideration and the corresponding bits in the Bloom vector are set.

To test for replay, when the next packet is received, the same is hashed using the *m* hash functions and the corresponding bits in the vector are tested. As before, if the bit values are 1, the packet is a replayed packet, otherwise it is not.

| Dest | AM | Len | Src | Ctr | Data | MAC |
|------|----|----|-----|-----|------|-----|
| (2) | (1) | (1) | (2) | (2) | (0...29) | (4) |

Fig. 4 TinySec_AE Packet Format [5]

| Dest | AM | Len | Data | MAC |
|------|----|----|------|-----|
| (2) | (1) | (1) | (0...29) | (4) |

Fig. 5 TinySec_Auth Packet Format [5]

The fact that false positives are possible here means that if a packet is not a replayed packet, it may be identified as a replayed one. The sender may in that case have to re-transmit the same packet. However, the converse, that is, even if the packet is a replayed packet and is flagged as a genuine one is never possible. Therefore, the overall semantics of the protocol remain sacrosanct.

Bloom filters are well suited in the severe resource-constrained environment of sensor nodes. This is so, because they scale well with the increase in the number of network nodes, unlike other techniques. Asymptotically, only $O(n \log n)$ bits are required to store $n$ messages. And, Bloom filters also exhibit the desirable property that adding and querying an element occurs in constant time.

## V. OUR DESIGN

In this section, we discuss our design of various replay protection schemes employed for the implementation and subsequent evaluation, in the TinySec link layer framework.

### A. Approach 1: Counter based algorithm

As mentioned earlier, using a counter value as the message tag is one of the most common approaches for detecting packet replays. In order to apply this technique, it is necessary to decide and identify a suitable data structure to maintain the value of the counter. For this purpose, we exploit to an advantage, the packet format of a TinySec packet itself shown in fig. 4.

For the *TinySec_AE* mode, in order to support both the encryption and message authentication, it is necessary to employ a block cipher mode of operation (e.g. the Cipher Block Chaining (CBC) mode [12]) and a message authentication code algorithm (e.g. CBC-MAC [11]). The Cipher Block Chaining (CBC) in turn needs an initialization vector (IV) to be defined with it, to ensure *semantic security* [5]. In TinySec packet format, the first four bytes of the packet make up the IV. The components of the IV are also shown in Fig. 4.

As can be observed, TinySec partitions the last 4 bytes of the IV into *src*||*ctr*, where *src* is the address of the sender and *ctr* is a 16-bit counter, used to ensure that the IV for every source node is distinct. Each time a sender sends message, the *ctr* value is incremented by 1.

We re-use the same counter value in our first algorithm, for ensuring replay protection, thereby obviating any reason for augmenting or changing the existing TinySec packet format.

In order to handle the possibility of the packets arriving out of order, at the destination, we employ a sliding-window. The window is maintained for each neighbor of the recipient node. If the value of the counter in received packet is less than that received in the previous packet, then it is checked within this window. If the same counter value is found within the window, then the packet is tagged as a replayed packet and is dropped. Otherwise it is accepted as a genuine packet and the counter value added to the window. The principal limitation is the polynomial increase in storage, with the increase in the number of neighbor nodes of a particular node. Hence, the scalability of this algorithm is limited.

However, this algorithm does not require counter resynchronization as in SNEP or in MiniSec unicast algorithm. The counter value required for replay protection is part of IV, so there is no extra overhead in packet size involved for this method. Hence, the method incurs very less consumption in radio energy.

It is essential to point out that that we have used this method with only the *TinySec_AE* mode that uses an IV (and a counter value therein) for the encryption operation. Thus, we are able to make use of the *Ctr* field because of the support available for IV. For those applications that demand confidentiality and message authentication both, our algorithm directly provides a new mode of operation with a new attribute i.e. replay protection with authentication and encryption.

However, for those applications that need only message integrity and hence employ the *TinySec_Auth* mode of operation, the IV field is not used; resulting into a different packet format as shown in fig. 5. Due to the absence of the *Ctr* value here, it would be necessary to modify our algorithm and make a provision for the *Ctr* field separately. Our basic algorithm can easily be extended for the purpose.

### B. Approach 2: Hash based algorithm

Unlike the previous algorithm, our second algorithm in implementing replay protection is TinySec-mode independent. Here, as the message type tag, we use a hash value generated by a hash function, instead of using a counter value. We have used SHA-1 [22] as the hash function. SHA-1 is a one-way hash function that produces 160-bit output, when message of any length less than 264 bits is input. The basic algorithm followed for replay detection is illustrated in the pseudocode in fig 6.

The recipient computes the hash value from the received packet using the hash function agreed upon into the variable *HashValue*. As in the previous case, in order to handle the packets delivered out of order, we maintain a window of appropriate size to store the previously received hash values.

On receiving the incoming packet, we first apply a hash on

```
Algorithm ReplayDetectHash(PacketReceived){
1.
HashValue=computeHashValue(PacketReceived);
2.  for id = 1 to lastValidId {
3.    if (id == DestIDInPacket) {
4.       for window = 1 to last {
5.          if (HashValue ==
                  HashValueInWindow[window])
6.                replayed = 1;
              else
7.                replayed = 0
8.       InsertHashValueinWindow();
       }
}
```

Fig. 6 Antireplay Algorithm `ReplayDetectHash`

it. The generated hash value is checked in the window of hash values, maintained for this neighbor. If the same hash value is found within the window, then the packet is considered as a replayed packet, otherwise it is considered as a genuine packet. For the genuine packets, the hash value is inserted into the window. In this method, the out-of-order packets therefore, are again automatically handled.

However, it is emphasized that only the packets replayed within the window will be detected as the replayed/non-replayed packets. Hence, the security of this scheme largely depends on the size of the window employed. Greater the size of the window, more precise the scheme is.

In addition, this scheme requires higher computational and storage resources, due to the complex hash computations and the need for storing 20-byte hash values. Therefore, it works only for small number of neighbors. However, with a light-weight hash function, this limitation can be overcome.

*C. Approach 3: Bloom filter based algorithm*

In the bloom filter based algorithm, therefore, we use a bloom filter to reduce the storage overhead. However, to implement a bloom filter *m* hash functions are required.

In the first version using the Bloom filter, we use only a single hash function. Each node *i* in the network is required to maintain a single bloom filter $BF_i$. For the sake of comparison, we employ SHA-1 as the hash function here, too.

The basic algorithm for replay detection using this method is as shown in Fig. 7.

```
Algorithm ReplayBloomFilter(Packet In, int n,
                         Filter BloomFilter)
{
1.  if (!FilterCreated) {
2.  Filter CreateBloomFilter(int n, Filter
                             BloomFilter);
3.    FilterCreated=1;
   }
4.  for each received packet {
5.    HashValue=ApplySHA1(TypeTag(Packet
                                     In))
6.    if (BloomFilter[HashValue] == 1)
7.         return (replayedpacket=1)
8.    else {
9.         BloomFilter[HashValue] = 1;
10.        return (replayedpacket=0);
      }
     }
}
```

Fig. 7 Antireplay Algorithm `ReplayBloomFilter`

On receiving an incoming packet, SHA-1 is applied on the whole packet and the bit corresponding to the resulting hash value is checked in the bloom filter. Again, the same bit-check paradigm, as before is employed. Thus, if the bit is not set, then the corresponding packet is treated as a non-replayed packet and is consequently accepted. Otherwise, it is tagged as a replayed packet.

As mentioned earlier, the memory requirement in such scheme is constant; being independent of the number of hash functions used as well as independent of the number of nodes used – thereby answering the scalability concerns. In addition, such an anti-replay policy would detect all the replayed packets, resulting in 0% false negative rate.

Considering the size of Bloom filter to be *n* and *k* hash functions, after inserting status of p packets in the filter, the probability [10] that a particular bit is still 0 is given by

$$P_0 = \left(1 - \frac{1}{n}\right)^{kp} = 1 - e^{\frac{-km}{n}} \qquad (5)$$

Therefore, the probability of a false positive i.e. the probability that all the k bits (for each hash function) have been previously set is given by [10]

$$p_{\text{false-positive}} = (1-p_0)^k = \left(1 - \left(1 - \frac{1}{n}\right)^{kp}\right) = \left(1 - e^{\frac{-km}{n}}\right)^k \qquad (6)$$

However in general, to prevent false positives, the bloom filter design criterion specifies tuning of the number of hash functions. That is, by selecting the size of the Bloom Filter and the number of hash functions, the probability of false positives can be lowered arbitrarily [10].

The false positive rate for a standard bloom filter employing *k* hash functions is approximately $1/2^k$ [23]. Hence, increasing the number of hash functions must result in the reduction of false positive rates. However, applying multiple hash functions results in increased consumption of CPU cycles and thereby energy.

In order to reduce this as much as possible, we use modified version of the previous algorithm, wherein, we use bloom filter with eight 32-bit hash functions i.e. theoretically $2^{32}$ different values per hash function. Thus, the false positive rate in our algorithm turns out to be 1/256 i.e. 0.00390.

```
Algorithm ReplayBloomFilter8(Packet In, Set
    HashFunction, int n, Filter BloomFilter)
{
1.  if (!FilterCreated) {
2.  Filter CreateBloomFilter(int n,
                  Filter BloomFilter);
3.    FilterCreated=1;
   }
4.  for each received packet {
5.    for i= 1 to k {
6.       HashValue=ApplyHash_i(TypeTag(Packet
                                      In))
7.    if (BloomFilter[HashValue] == 1)
8.       return (replayedpacket=1)
      else {
9.       BloomFilter[HashValue] = 1;
10.      return (replayedpacket=0);
      }
     }
    }
}
```

Fig.8 Antireplay Algorithm `ReplayBloomFilter8`

With Bloom filter of length *n*, a set of *k* hash functions each of size *m* bits, for each message received at the destination, the antireplay algorithm is as shown in Fig. 8.

The eight hash functions that we have employed are from the family of universal hash functions, requiring lesser computational overhead than SHA-1 viz. *RSHash, JSHash, PJWHash, ELPHash, BKDRHash, SDBMHash, DJBHash, DEKHash*, and *APHash*.

We carefully implement the bloom filter as two-dimensional vector in order to feasibly implement it.

We must emphasize here that for any algorithmic solution, the scalability and the impending overhead concerns must be dealt with, at the algorithm design level itself. Such concerns must not be left out to be tackled in the implementation level. As compared to the counter- based algorithm or even the SHA-1 hash function based algorithm, we follow the later dictum in our Bloom filter based algorithm. The fact becomes clearer and tangible with our experimentation results analyzed in section VII.

## VI. METHODOLOGY AND TOOLS USED

In this section, we describe the platforms and tools used in our experimentation and the test application employed.

### A. Evaluation Platform & Tools

We have implemented the replay protection schemes described, for the Mica2 motes. We have used TinyOS version 1.x as the operating system [25] having the integrated support for the TinySec library. However, in order to incorporate replay protection algorithms, we made appropriate changes to the TinySec link layer security framework. This includes creation of new modules and interfaces as well as modifications in system files of TinyOS.

We implemented our code in nesC [26], the programming language used for TinyOS. In addition, to simulate and test our implementation, we used the TOSSIM [27] simulator that runs on an Intel x86 platform.

TOSSIM does not support energy and CPU cycles simulations. So, for energy and CPU cycle analysis, we use Avrora [28], an instruction level event simulator. Using results obtained from TOSSIM and Avrora, we evaluate the performance of our replay protection algorithms.

### B. Methodology Adopted

We use a simple application that comes bundled with the TinySec environment, to test our algorithms. The call-graph of the application is as shown in figure 9.

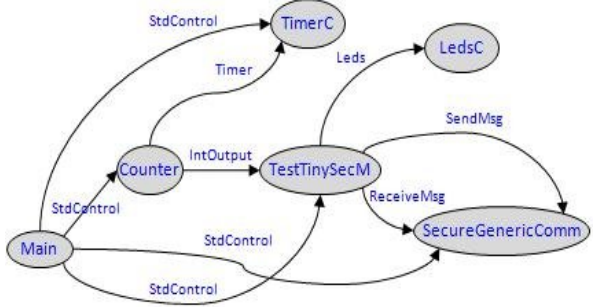

Fig. 9. The TestTinySec Call-graph

```
Algorithm TestTinySec
1. counter = timer
2. while (counter == fired){
3.   if (Send(Data Packet)) then
                           LED=green
4.    else if( Receive(Data Packet))
      then LED=red
  }
```

Fig. 10. The TestTinySec Algorithm

The pseudocode of our test application is shown fig. 10.

As can be observed, the *TestTinySecM* module is the main component implementing the application. *TestTinySecM* implements a counter that is incremented on firing of the timer. The counter value modified by the component *Counter*, is further passed by *TestTinySecM* through the *SendMsg* interface for onward transmission over the radio, to the component *SecureGenericComm*. In addition, when the message is sent, the *Leds* interface is used to toggle the LED green when the message is transmitted by a mote, whereas to turn it red, when the message is received by a mote. All the messages are encrypted and authenticated over the air.

A partial call-graph of the same application with the security modules into play is shown in fig. 11. This call-graph was generated with the default crypto modules of TinySec – viz. the Skipjack cipher [13] wired in the Cipher Block Chaining (CBC) mode for encryption [12] and the Cipher Block Chaining Message Authentication Mode (CBC-MAC) [11] for message integrity.

The main module of the TinySec library is *TinySecM* that is responsible for invoking all the security related commands.

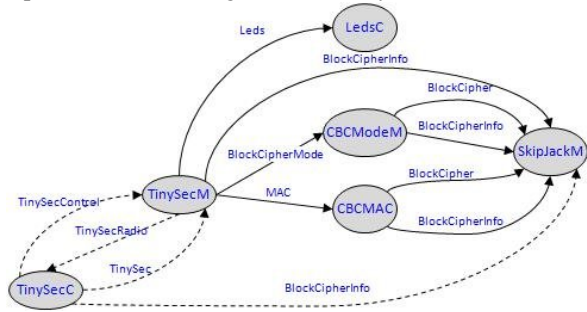

Fig. 11. The Test Application with Security Components into play

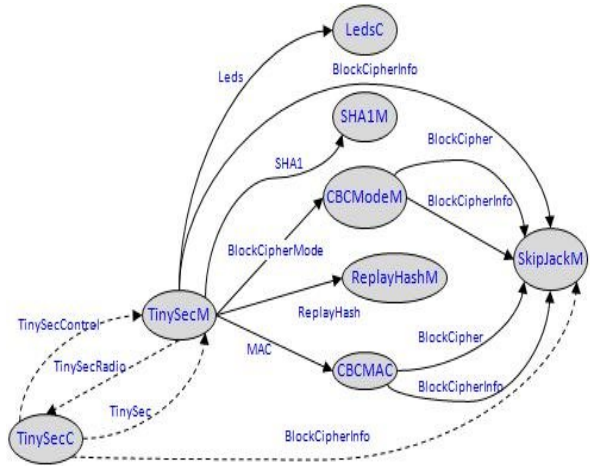

Fig. 12. Our implementation of the Test Application with SHA-1 function

*TinySecC* is the configuration file that shows wiring between different modules. TinySec does not provide replay protection; hence there is no module providing the functionality. Next, we show component graph of our second algorithm i.e. *Hash(SHA-1)* based algorithm in figure 12. For other algorithms, similar graph can be generated with the appropriate change in the name of the module as per the use. As shown, we have implemented *SHA-1M* module which provides the interface *SHA-1* and is responsible for calculating the hash on the message.

We have also implemented *ReplayHashM* module which provides *ReplayHash* interface. We modify the TinySec configuration files so that *TinySecM* can use these interfaces viz. *SHA-1* and *ReplayHash* and call the commands implemented therein, to provide replay protection.

VII. EXPERMIENTAL RESULTS & ANALYSIS

We have evaluated the replay protection algorithms based on different metrics viz. memory, energy and CPU cycles. In this section, we show our experimental results for these algorithms. Based on these results, we propose the suitable candidate for implementing replay protection.

A. *Storage Requirements*

We have used the two parameters viz. *Number_of_Neighbors* and *Window_Size* to evaluate the storage requirements of our schemes for Counter-based algorithm and Hash-based algorithm. In figure 13, we show the RAM requirements for these two algorithms with varying number of neighbors (for a fixed window size of 8).

The graph is almost linear as increasing the number of neighbors, increases the storage requirements, for maintaining the state of previously received counter values. We maintain the state in 2-byte counter in our counter based algorithm. In SHA-1 based algorithm we do so in a 20-byte message digest for all the neighbors.

Hence it is clear that the memory requirement for SHA-1 based algorithm is higher than that for the counter based algorithm. We experienced that the counter based algorithm works for a maximum of 150 motes, whereas the SHA-1 based algorithm works for maximum 7 motes with fixed window size of 8.

In figure 14, we show the RAM requirements for the counter based and SHA-1 based algorithms with varying window size, but the fixed numbers of neighbors at 5. We can see from the figure that the counter based algorithm works for maximum window size up to 310, whereas SHA-1 based algorithm works for maximum window size set to 13.

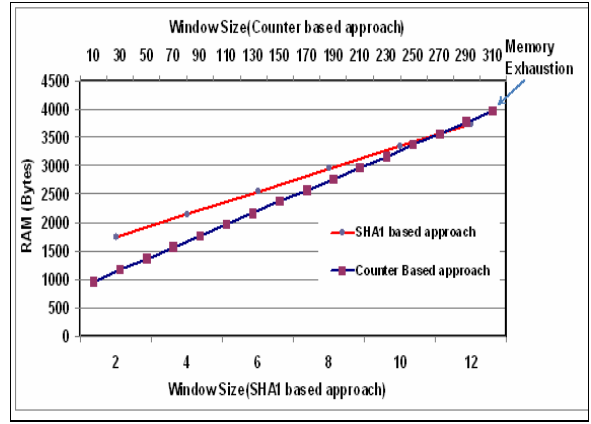

Fig. 14. RAM requirements for Counter and Hash-based algorithms with varying window-size for a fixed number of neighbors

In a congested network, the bigger window size will result in good performance. Small window size will fail to detect most of the replayed packets as the previously received packets are continuously being replaced by newly arriving packets.

However, for the applications with significantly lower packet transmission rate, smaller window size can be appropriate. For such applications, our scheme works well for dense network also.

It is clear from our results that increasing the number of neighbors (hence the number of motes) results in increased demand for RAM for Counter-based and SHA-1-based algorithms. In figure 15, we portray the RAM requirements for all of our algorithms including the Bloom-filter based, too. For Bloom filter based algorithms, RAM requirements remain constant irrespective of number of neighbors.

We summarize the results of RAM usage of all the algorithms in Table I. We use the TinySec without replay protection as the basis for comparing the percentage increase in RAM usage for all the algorithms in a network with 10 nodes.

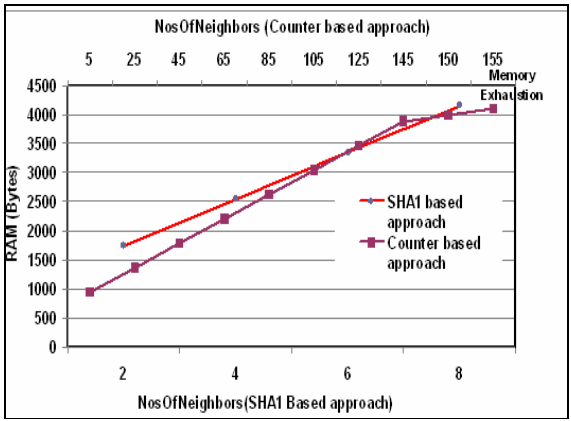

Fig. 13. RAM requirements for Counter and Hash-based algorithms with varying number of neighbors for a fixed window-size of 8

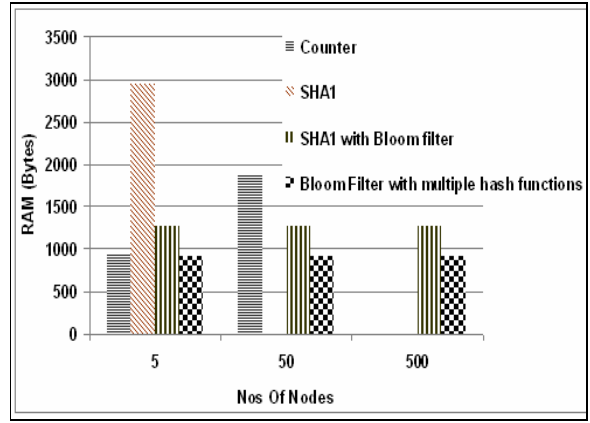

Fig. 15. RAM requirements for all of the algorithms

Table I. Percentage Increase in RAM for all algorithms

| Method Description | RAM usage (Bytes) | % increase over TinySec |
|---|---|---|
| TinySec without replay protection | 840 | - |
| Counter Based algorithm | 1050 | 25 |
| Hash based algorithm | 4958 | 489 |
| Bloom filter with single hash function | 1258 | 49.7 |
| Bloom filter with multiple hash functions | 904 | 7.62 |

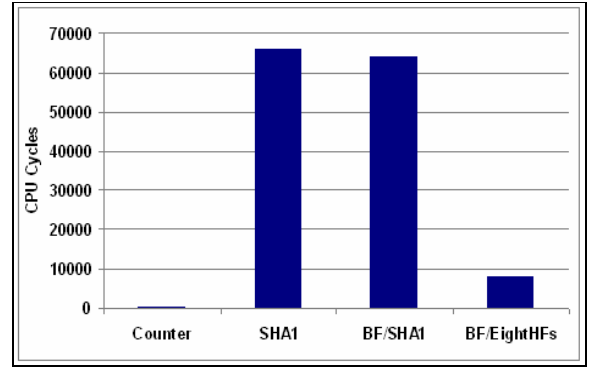

Fig. 17. The CPU cycles usage for all algorithms under evaluation

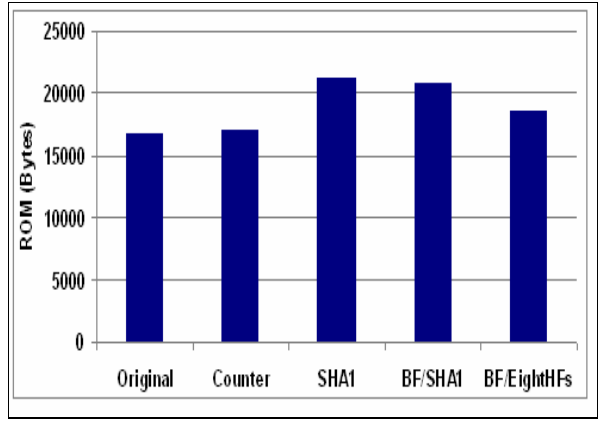

Fig. 16. ROM for all of the algorithms under evaluation

It is clear from Table I that Bloom filter based algorithm with multiple hash functions results in less than 10% overhead in memory, at the same time giving good performance.

This statistic is even more significant than it appears because, the advancement in mote technology has shown higher growth rate in ROM availability than that of RAM.

Hence, even with the newer generation motes, RAM remains a scarce resource. Therefore, an algorithm that uses fewer RAM is definitely better.

We show the results of ROM usage in figure 16. The BF(SHA1) and BF(8 HFs) depict the results for Bloom filter with one hash function and that with 8 hash functions respectively. The SHA-1 based algorithms require marginally higher ROM than other algorithms because of the complex computations involved in SHA-1.

### B. CPU cycles usage

The results obtained for the CPU cycle requirement of our implementation, using Avrora are as shown in figure 17. The computational overhead for SHA-1 results in higher number of CPU cycles required for SHA-1 based algorithms.

Sensor nodes are deployed in hostile, unattended environment. Here, it might not be possible to replenish the energy resources and hence, power consumption is a critical aspect while designing the sensor network algorithms.

Power consumption can be minimized by ensuring that the CPU cycles consumed are as less as possible. In our implementation, we do not require extra overhead in packet size. Hence, no extra overhead incurred for transmitting additional information.

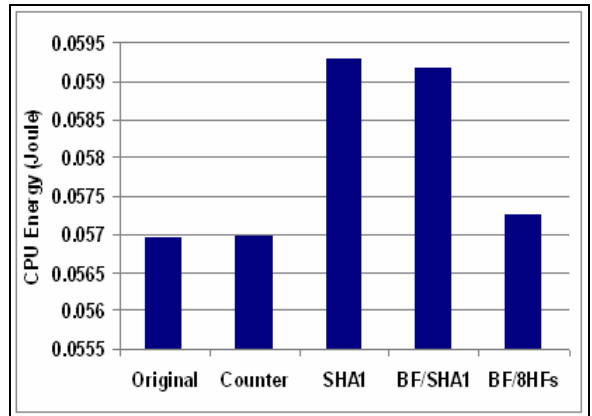

Fig. 18. Energy requirements for all algorithms under evaluation

### C. Energy Requirements

The results for the energy requirements of our implementation are shown in figure 18. The energy overhead incurred is basically for execution of replay related operations only which is considerably less as compared to packet overhead.

In Table II, we show the percentage increase in the energy overhead for all algorithms over TinySec without replay protection.

It is clear from our results that our first and last algorithms incur just 0.06% and 0.17% overhead in energy. As mentioned earlier, the counter based algorithm lacks scalability. Hence, the bloom filter based algorithm offers the best of both worlds. The SHA-1 based algorithm entails higher overhead, because of its higher computational requirements, as mentioned before.

### D. Analysis

Our experimental results clearly demonstrate that counter based and SHA-1 based algorithms work only for a limited number of nodes in the network due to the limited memory of the sensor nodes. For dense network, these two algorithms do not give good results. In addition performance depends greatly on window size.

For larger window sizes, these schemes give good result at the same time demanding higher RAM. Hence, there is a tradeoff between performance and RAM.

Table II. Percentage Increase in Energy Consumption for all algorithms

| Method Description | % increase in energy over TinySec |
|---|---|
| TinySec without replay protection | - |
| Counter Based algorithm | 0.06 % |
| Hash based algorithm | 4.14 % |
| Bloom filter with single hash function | 3.91 % |
| Bloom filter with multiple hash functions | 0.17 % |

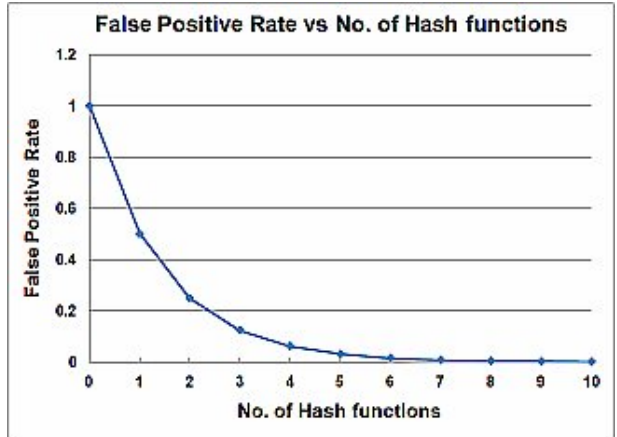

Fig. 19. Effect of hash functions on false positives

RAM requirements for bloom filter based algorithm are constant for any no. of nodes. As shown in Table I, our experimental results with 10 nodes, clearly depict that with less than 10% increase in RAM, replay protection can feasibly be implemented in link layer security architecture.

Bloom filter based algorithm works for any arbitrary network. It guarantees 0% false negatives but generates false positives. The rate of false positives increases as large number of packets is inserted in the filter. Increasing the number of hash functions will result in less false positives. We have used five hash functions in our algorithm.

We depict the relation between the false positive rate and the number of hash functions used in figure 19. As can be seen, for eight hash functions, the false positive rate is less than 0.5. Because of this reason, our last algorithm viz. bloom filter with multiple hash functions is the preferred algorithm, giving better throughput at the same time keeping memory requirement constant.

## VIII. CONCLUSION

In this paper, we propose a novel yet simple algorithm to ensuring replay protection at the link layer framework in the WSNs. We also justify the need of replay protection at the link layer framework only in WSNs, discuss in detail the common algorithms for ensuring replay protection in conventional as well as the WSNs and evaluate the algorithms against the novel Bloom filter based algorithm that we propose.

Our experimental results clearly demonstrate that bloom filter based algorithm with multiple hash function works well with lesser than 0.2% increase in energy and less than 10% increase in memory as compared to the one without replay protection enabled.

This work can further be extended with the design and implementation of a flexible link layer security framework in which as one of the security attribute, our implementation and analysis here can be used. We propose the basic design of such architecture in [29] that is principally motivated from the diverse security requirements of the broad spectrum of WSN applications [30]. Such architecture must not only be flexible with respect to the security attributes but also the radio bandwidth and transceiver chip, motes with a support for the next generation motes etc. Our current work is focused on implementing the same.

ACKNOWLEDGMENT

The authors are grateful to all the anonymous reviewers for taking pains to devote their time for reviewing this paper and for the useful suggestions.

**Devesh C. Jinwala** was born on 3[rd] July 1964. He has a Master's degree in Electrical Engineering from the Maharaja Sayajirao University of Baroda, India with specialization in Microprocessor Systems and Applications.

He is employed as an Assistant Professor in Computer Engineering with Sardar Vallabhbhai National Institute of Technology, Surat (India) since 1991. He is currently working on Configurable Link layer Security Protocols for Wireless Sensor Networks. He has numerous publications in the area including those in the Lecture Notes in Computer Science Series (Springer Verlag, Heidelberg, Germany). His major areas of interest are Information Security Issues in Resource Constrained Environments, Algorithms & Computational Complexity and Software Reliability, Verification & Testing.

**Dhiren R. Patel** was born on 29[th] July 1966. He has a Master's degree in Computer Science & Engineering from IIT Kanpur, India and Ph D in Computer Engineering from the South Gujarat University (REC Surat), India.

He is employed as a Professor of Computer Engineering at NIT Surat, India. His major areas of interest are Information/Network Security, Web Engineering and Ubiquitous Architectures. Apart from his numerous International publications and distinguished talks, He has also authored a book "Information Security: Theory & Practice" published by Prentice Hall of India in 2008

**Sankita Patel** was born on 20[th] Dec 1980. She has a Masters Degree and a Bachelors Degree in Computer Science and Engineering from S. V. National Institute of Technology, Surat.

She is currently employed as an Assistant Professor at NIT Surat, India. Her major areas of interest are Information Security & Privacy in Data Mining.

**Kankar S. Dasgupta** has a Masters Degree in Computer Science and Engineering from Jadavpur University, Kolkata and a Doctorate in Electrical Engineering from the Indian Institute of Technology, Bombay.

He is currently the Director of DECU at the Indian Space Research Organization, Ahmedabad.